\documentstyle[prl,aps,preprint,epsf, epsfig]{revtex}

\begin{document}
\tighten
\title{Probing the Nucleon's Transversity Via Two-Meson Production
in Polarized Nucleon-Nucleon
 Collisions\thanks{This work is supported in part by funds 
provided by the U.S.
Department of Energy (D.O.E.) under cooperative 
research agreement \#DF-FC02-94ER40818.}}

\author{Jian~Tang}

\address{{~}Center for Theoretical Physics\\ 
Laboratory for Nuclear Science\\ 
and Department of Physics \\
Massachusetts Institute of Technology\\
Cambridge, Massachusetts 02139 \\
{~}}

\date{MIT-CTP-2769 ~~~ hep-ph/xxxxxxx ~~~
Submitted to {\it Physical Review D} ~~~ July 1998}

\maketitle

\begin{abstract}

We explore the possibility of probing the nucleon's transversity 
distribution $\delta q(x)$
through the final state interaction between two mesons 
($\pi^+\pi^-$, $\pi K$, or 
$K\overline K$)
produced  in transversely polarized nucleon-nucleon collisions.
We present a single spin asymmetry and estimate its magnitude 
under some assumptions for the transversity distribution 
function and the unknown
interference fragmentation function. 

\end{abstract}
 

\narrowtext

The quark transversity distribution  in the nucleon $\delta q(x)$
measures the probability difference to
find a quark polarized along versus opposite to  the polarization of a
nucleon polarized transversely to its direction of motion
\cite{ralston79,jaffe91,artru93,cortes92}. 
Along with unpolarized and 
longitudinally polarized quark distributions, it completely 
characterizes the state of quarks
in the nucleon at leading twist in high-energy processes. 
While the other two have been studied extensively in the past
through various high-energy experiments, very little is known about
the transversity distribution $\delta q(x)$ 
since it decouples from hard QCD processes
at the leading twist due to its 
chiral-odd property. For example, it is suppressed like 
${\cal O}(m_q/Q)$ in totally-inclusive deep inelastic scattering (DIS). 

In our recent works\cite{jjt1}, we have studied the semi-inclusive
production of two mesons ({\it e.g.\/} $\pi^+\pi^-$, $\pi K$, or 
$K\overline K$) in the current fragmentation region in 
DIS on a transversly polarized nucleon. We have shown that 
the interference effect between the $s-$ and $p-$wave of the 
two-meson system around the $\rho$ (for pions), 
$K^*$ (for $\pi K$), or $\phi$ (for kaons) provides 
a single spin asymmetry which may be sensitive to the 
quark transversity distribution in the nucleon. 
Such interference allows the  quark's
polarization information to be  carried through 
the quantity $\vec k_+
\times \vec k_- \cdot  \vec S_\perp$,  where
$\vec k_+$, $\vec k_-$, and $\vec S_\perp$ are the
three-momenta of 
$\pi^+$ ($K$), $\pi^-$ ($\overline K$), and the nucleon's
transverse spin, respectively. This effect appears at the leading twist
level, and the production  rates for pions and kaons are large in
DIS.  However, it would vanish by
T-invariance in the  absence of final state interactions, or by
C-invariance if the two-meson state were an eigenstate of
C-parity. Hence there is no effect in the regions of the two-meson 
mass dominated by a single resonance.
However,  both suppressions are evaded in the $\rho$
($\pi^+\pi^-$), $K^*$ ($\pi K$), and 
$\phi$ ($K\overline K$) mass regions where 
both $s-$ and $p-$wave production channels are active.

In this paper, we extend our study to discuss the possibility of probing
the quark transversity distribution in the nucleon
via two-meson semi-inclusive production in transversely polarized 
nucleon-nucleon collisions.  
Various processes in transversely polarized nucleon-nucleon
collisions have been suggested to measure the nucleon's transversity
distribution since it was first introduced about
two decades ago\cite{ralston79}, among which are 
transversely polarized Drell-Yan \cite{ralston79} and
two-jet production\cite{ji92,jaffe-saito}. However,
the Drell-Yan cross section is small and requires an  
antiquark transversity distribution, which is likely to be quite small.
The asymmetry obtained in two-jet production is rather small
due to the lack of a gluon contribution\cite{ji92,jaffe-saito}.
On the other hand, in the process described here, 
the gluon-quark scattering dominates and only
one beam need be transversely polarized.
Unless the novel interference fragmentation function is
anomalously small, this will provide a feasible way to
probe the nucleon's transversity distribution. 
The results of our analysis are summarized by Eq.~(\ref{asymmetry})
where we present the asymmetry for $\pi^+\pi^- (\pi K, K\overline{K})$
production in transversely polarized 
nucleon-nucleon collisions. 

\begin{figure}[h]
\begin{minipage}[h]{6.0in}
\epsfxsize=3.0in
\centerline{\epsffile{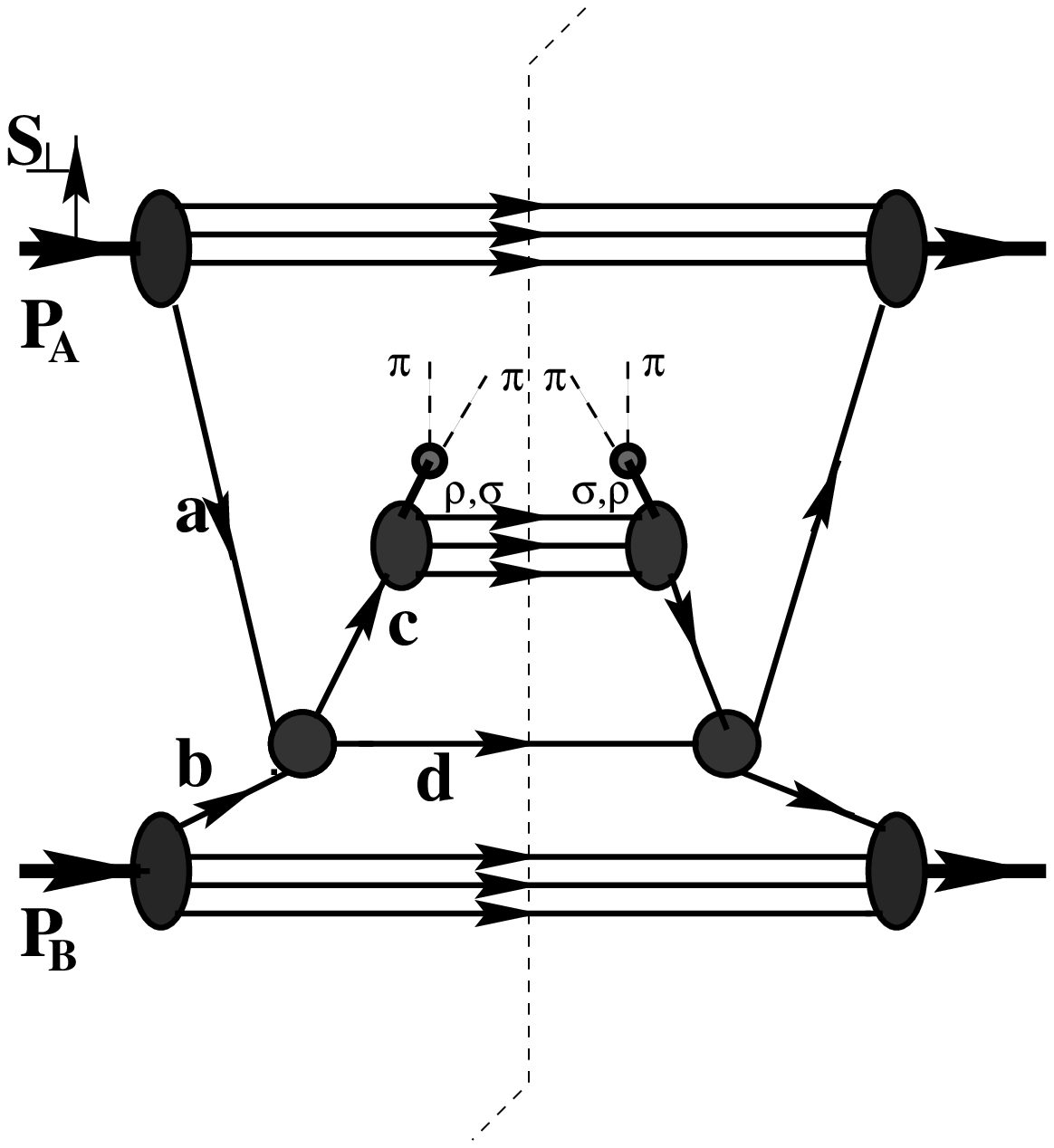}}
\caption{Hard scattering diagram for two-meson semi-inclusive production 
in nucleon-nucleon collision.}
\label{fig1}
\end{minipage}
\end{figure}

We consider the semi-inclusive nucleon-nucleon collision process with 
two-pion final states being detected: $N\vec N_\bot\rightarrow 
\pi^+\pi^- X$. (The analysis to follow applies as well to 
$\pi K$ or $K\overline{K}$ production.) One of the nucleon 
beams is transversely polarized with polarization vector $S_\mu$, and
momentum $P_\mu^A$. 
The other is unpolarized, with momentum denoted by $P_\mu^B$.
The experimentally observable invariant variables are defined as  
$s\equiv (P_A + P_B)^2,\hspace*{0.5cm}t\equiv (P_h - P_A)^2,
\hspace*{0.5cm} u\equiv (P_h-P_B)^2$,  and 
the invariants for the underlying partonic processes are 
$\hat{s}\equiv (p_a + p_b)^2,\hspace*{0.5cm} \hat{t}
\equiv (p_c - p_a)^2,
\hspace*{0.5cm} \hat{u}\equiv (p_c-p_b)^2$,
where $P_h$ is the total momentum of the two-pion system,
$p_a$, $p_b$, $p_c$, $p_d$ are the momenta for the underlying
partonic scattering processes (see Fig.~\ref{fig1}).
The longitudinal momentum fractions $x_a$, $x_b$ and $z$ are
given by $p_a=x_aP_A$, $p_b=x_bP_B$ and $P_h=zp_c $.
The $\sigma [(\pi\pi)^{I=0}_{l=0}]$ and $\rho [(\pi\pi)^{I=1}_{l=1}]$
resonances are produced with momentum
$P_h$. We recognize that the
$\pi\pi$ $s$-wave is not resonant in the vicinity of the $\rho$ and our
analysis does not depend on a resonance approximation.  For simplicity we
refer to the non-resonant $s$-wave as the ``$\sigma$''.  

The invariant squared mass of the two-pion system is 
$m^2 = (k_++k_-)^2$, with $k_+$ and $k_-$ the momentum of
$\pi^+$ and 
$\pi^-$, respectively. The decay polar angle in the rest frame of the
two-meson system is denoted by $\Theta$, and the azimuthal angle $\phi$ is
defined as the angle of the normal of two-pion plane with respect to the
polarization  vector $\vec S_\perp$ of the nucleon, 
$\cos\phi = {{\vec k_+}\times{\vec k_-}\cdot\vec S_\perp / |\vec k_+\times
\vec k_-||\vec S_\perp|}$. This is the analog of the ``Collins angle'' defined 
by the $\pi^+\pi^-$ system \cite{collins94}.

Since we are only interested in a result at the leading twist, we follow the
helicity density matrix formalism developed in Refs.~\cite{jaffe95,jaffe96},
in which all spin dependence is summarized  in a {\it double} helicity
density matrix. We factor the process at hand 
into basic ingredients (See Fig.~\ref{fig1}): 
the $N\rightarrow q$ (or $N\rightarrow g$) 
distribution
function, the hard partonic 
$q_aq_b\rightarrow q_cq_d$ cross section,  the $q \rightarrow (\sigma, \rho)$
fragmentation, and the decay $(\sigma, \rho)\rightarrow \pi^+\pi^-$, all as
density matrices in helicity basis: 
\begin{eqnarray}
\left[{{d^7\sigma(N\vec N_\perp\rightarrow \pi^+\pi^- X)}
\over{dx_a\, dx_b\, d\hat{t}\, dz\, dm^2\, d\cos\Theta\,
d\phi}}\right]_{H'H}&&
\nonumber
\\*[14.4pt]
&&\hspace*{-2.8cm}=\left[{\cal F}(x_a)\otimes {{d^3\sigma(q_a
q_b\rightarrow q_cq_d)}
\over{dx_a\, dx_b\, d\hat{t}}}\otimes
{{d^2\hat{\cal M}}\over{dz\,
dm^2}}\otimes
{{d^2{\cal D}}\over{d\cos\Theta\, d\phi}}\otimes {\cal F}(x_b)\right]_{H'H}
\end{eqnarray}
where $H(H')$ are indices labeling the helicity states of
 the polarized nucleon. In order
to incorporate the final state interaction,
we have separated the $q\rightarrow \pi^+\pi^-$ fragmentation process into 
two steps. First, the quark fragments into the resonance ($\sigma$,
$\rho$) ,  then the resonance decays into two pions, as shown in the middle
of the Fig.~\ref{fig1}.     

The $s-p$ interference fragmentation functions describe the emission
of a $\rho(\sigma)$ from a parton, followed by absorption of 
$\sigma(\rho)$ forming a parton. Imposing various symmetry 
(helicity, parity and
time-reversal) restrictions, the interference fragmentation 
can be cast into a double density matrix
notation\cite{jjt1}
\begin{eqnarray}
{d^2 \hat{\cal M}\over dz\, dm^2}
= &&
\Delta_0(m^2)\left\{I\otimes \bar\eta_0\,
\hat{q}_{_I}(z)
+\left(\sigma_+\otimes \bar\eta_-
+ \sigma_-\otimes \bar\eta_+\right)\delta\hat{q}_{_I}(z)\right\}
\Delta^*_1(m^2)
\nonumber
\\
& &
+\Delta_1(m^2)\left\{I\otimes \eta_0\,
\hat{q}_{_I}(z)
+\left(\sigma_-\otimes \eta_+ 
+ \sigma_+\otimes\eta_-\right)\delta\hat{q}_{_I}(z)
\right\}\Delta^*_0(m^2)\ ,
\label{fragmentation}
\end{eqnarray}
where $\sigma_\pm\equiv (\sigma_1\pm i\sigma_2)/2$ with
$\{\sigma_i\}$ the usual Pauli matrices. The $\eta$'s are $4\times 4$ matrices
in $(\sigma, \rho)$  helicity space with nonzero elements only in the first
column, and the $\bar\eta$'s are the transpose matrices 
 ($\bar\eta_0 = \eta_0^T, \bar\eta_+=\eta_-^T,
\bar\eta_-=\eta_+^T$), with the first rows $(0,0,1,0)$,
$(0,0,0,1)$, and $(0,1,0,0)$ for $\bar\eta_0$, $\bar\eta_+$, and $\bar\eta_-$,
respectively. The explicit definition
of the fragmentation functions will be given in Ref.~\cite{jjt3}.

The final state interactions between the two pions are  included
explicitly in
\begin{equation}
\Delta_0(m^2)=-i \sin\delta_0
e^{i\delta_0}\ ,\hspace*{1cm} 
\Delta_1(m^2)=-i \sin\delta_1
e^{i\delta_1}\ ,
\label{propagators}
\end{equation}
where $\delta_0$ and $\delta_1$ are the strong interaction $\pi\pi$ phase
shifts which can be determined by the
$\pi\pi\: {\cal T}$-matrix \cite{scattheory}.
Here we have suppressed the $m^2$ dependence  of the phase shifts for 
simplicity. 

The decay process,
$(\sigma,\rho)\rightarrow \pi^+\pi^-$,  can be easily calculated and encoded into
the helicity matrix formalism. The result for the interference part is
\cite{jjt1}
\begin{equation}
{d^2 {\cal D}\over d\cos\Theta\, d\phi}
={\sqrt{6}\over 8\pi^2 m}
\, \sin\Theta \left[ie^{-i\phi}\left(\eta_--\bar\eta_-\right)
+ie^{i\phi}\left(\eta_+-\bar\eta_+\right)
-\sqrt{2}\cot\Theta \left(\bar\eta_0 +\eta_0\right)\right]\ .
\end{equation}
Here we have adopted the customary conventions for the $\rho$ polarization
vectors, $\vec{\epsilon}_{\pm} = {\mp}(\hat{e}_1\pm i\hat{e}_2)/\sqrt{2}$
and $\vec{\epsilon}_0=\hat{e}_3$ in its rest frame with 
$\hat e_j$'s the unit vectors.

In the double density matrix notation, the quark distribution
function ${\cal F}_q(x)$ in the nucleon can be expressed as \cite{jaffe96}
\begin{equation}
{\cal F}_q(x) = {1\over 2} q(x)~I\otimes I + {1\over 2} \Delta
q(x)~\sigma_3 \otimes 
\sigma_3+{1\over 2}  \delta q(x)~
\left(\sigma_+\otimes\sigma_-+\sigma_-\otimes\sigma_+\right)\ ,
\label{calf}
\end{equation}
where the first matrix in the direct product is in the
nucleon helicity space and the second in the quark helicity space.
Here $q(x)$, $\Delta q(x)$, and $\delta q(x)$ are the spin average, helicity
difference, and transversity distribution functions, respectively,
and their dependences on $Q^2$ have been suppressed.

The gluon distribution function ${\cal F}_g(x)$ in the nucleon 
can be written as 
\begin{equation}
{\cal F}_g(x) = {1\over 3} G(x)~I\otimes I_g + {1\over 3} \Delta
G(x)~\sigma_3 \otimes 
S_g^3\ ,
\label{calg}
\end{equation}
where $I_g$ and $S_g^3$ are $3\times 3$ matrices in gluon helicity space
with nonzero elements only on the diagonal: diag($I_g$)=$\{1,1,1\}$
and diag($S_g^3$)=$\{1,0,-1\}$ .
Here $G(x)$ and $\Delta G(x)$ are the spin average and helicity
difference gluon distributions in the nucleon, respectively,
and, just like in Eq.~(\ref{calf}),
their dependences on $Q^2$ have been suppressed.
Note that there is no gluon transversity distribution
$\delta G(x)$ in the nucleon at the leading twist 
due to helicity conservation.
This is one of the reasons why transverse asymmetries in two-jet
production are typically small, as pointed out by Ji\cite{ji92},
Jaffe and Saito\cite{jaffe-saito}.

Several hard partonic processes contribute here, as
shown in the middle of Fig.~\ref{fig1}. The cross sections can be 
written as follows (here we list only the relevant parts, i.e. 
spin-average and transversity-dependent ones),
\begin{equation}
{d^3\sigma(q_aq_b\rightarrow q_cq_d)\over dx_a\, dx_b d\hat{t}}= 
{{\pi\alpha_s^2}\over{2\hat{s}^2}}\, 
I_b\otimes\left[{\hat{\bar{\sigma}}}_{ab}^{cd}\, I_a\otimes I_c  
  + 4 \delta\hat{\sigma}_{ab}^{cd}\, 
\left(\sigma_a^+\otimes\sigma_c^-+\sigma_c^-\otimes\sigma_a^+\right)\right]\ ,
\label{sigmahel}
\end{equation}
where subscripts $a$, $b$, $c$ means that the helicity matrices above are
in $a$, $b$, $c$ parton helicity spaces, respectively (See Fig.~\ref{fig1}).
 $\hat{\bar{\sigma}}_{ab}^{cd}$ and 
$\delta\hat{\sigma}_{ab}^{cd}$ are the spin-average and 
transversity-dependent cross sections for the underlying 
partonic processes $q_aq_b\rightarrow q_cq_d$, respectively,
which are shown in the 
Table I. 

\begin{figure}[h]
\begin{minipage}[h]{6.0in}
\vspace*{2.0cm}
\epsfxsize=3.5in
\centerline{\epsffile{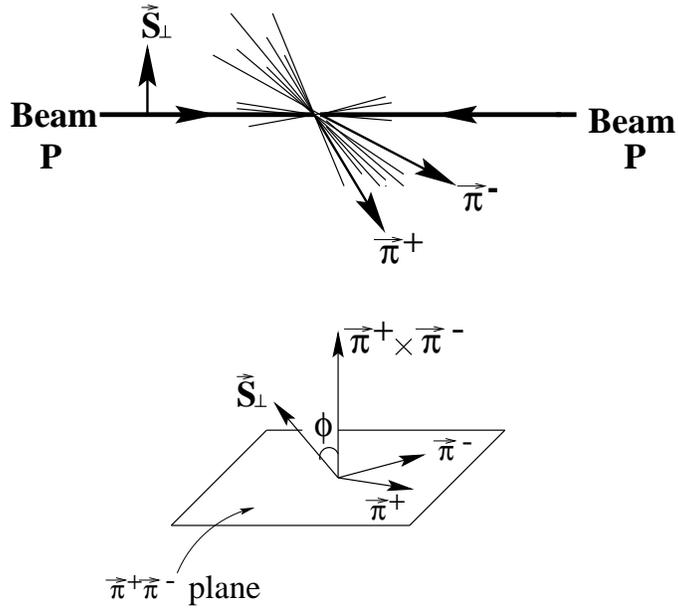}}
\end{minipage}
\vspace*{1.0cm}
       \caption{
           Illustration of the $pp$ collision at the center-of-mass frame
           and the so-called ``Collins angle'' $\phi$.
}
        \label{fig4}
\end{figure}
\begin{figure}[h]
\begin{minipage}[h]{6.0in}
\vspace*{-1.0cm}
\epsfxsize=3.0in
\centerline{\epsffile{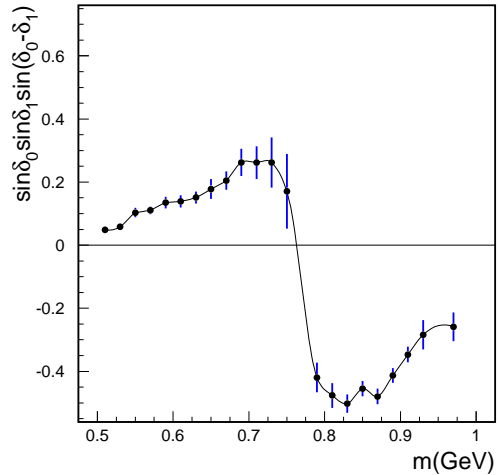}}
\end{minipage}
       \caption{
           The factor, $\sin\delta_0 \sin\delta_1
\sin(\delta_0-\delta_1)$, 
           as a function of the invariant mass $m$ of two-pion
system. 
           The data on $\pi\pi$ phase shifts are taken from
Ref.~\protect\cite{martin74}.
}
        \label{fig2}
\end{figure}

Combining all the above ingredients together, and integrating over $\Theta$ to 
eliminate the $\hat q_I$ dependence,  
we obtain a single spin asymmetry as follows,

\begin{eqnarray}
{\cal A}_{\bot\top}&\equiv& {d\sigma_\bot -d\sigma_\top
\over d\sigma_\bot +d\sigma_\top}
=-\frac {\sqrt{6}\pi}{4}\sin\delta_0\sin\delta_1\sin(\delta_0-\delta_1)
\cos\phi\nonumber \\
&&\otimes {{[\delta q(x_a)\otimes G(x_b)\otimes\delta\hat{q}_I(z)]
\delta\hat{\sigma}_{qg}^{qg}
+[\delta\bar{q}(x_a)\otimes G(x_b)\otimes\delta\hat{\bar{q}}_I(z)]
\delta\hat{\sigma}_{\bar{q}g}^{\bar{q}g}
+...
}\over{\{[G(x_a)\otimes G(x_b)]\hat{\bar\sigma}_{gg}^{q\bar q}+
[q(x_a)\otimes G(x_b)]
\hat{\bar\sigma}_{qg}^{qg}
+...\}\otimes[\sin^2\delta_0\hat{q}_0(z)+
\sin^2\delta_1\hat{q}_1(z)]}}
\label{asymmetry}
\end{eqnarray}
where $\hat{q}_0(z)$ and $\hat{q}_1(z)$ are 
spin-average fragmentation functions
for the $\sigma$ and $\rho$ resonances, respectively, and 
the summation over flavor is suppressed for simplicity. 
The terms denoted by $...$ include quark quark and quark antiquark
scattering contributions. 
This asymmetry can be
measured either by flipping the target transverse spin or by binning events
according to the sign of the crucial azimuthal angle $\phi$ 
(See Fig.~\ref{fig4}). The ``figure of merit'' for
this asymmetry, $\sin\delta_0\sin\delta_1\sin(\delta_0-\delta_1)$,
is shown in Fig.~\ref{fig2}.

The flavor content of the asymmetry ${\cal A}_{\bot\top}$ can be
revealed by using isospin symmetry and charge conjugation
restrictions.  For $\pi^+\pi^-$ production, isospin symmetry gives
$\delta\hat{u}_{I} = - \delta\hat{d}_{I}$ and $\delta\hat{s}_{I} =
0$.  Charge conjugation implies $\delta \hat{q}^a_{I}=-\delta
\hat{\bar q}^a_{I}$.  Thus there is only one independent interference
fragmentation function for $\pi^+\pi^-$ production, and it may be factored
out of the asymmetry, e.g. $\sum_a \delta q_a\delta\hat{q}_{I}^a=
[ (\delta u - \delta\bar u)- (\delta d - \delta\bar
d)]\delta\hat{u}_{I}$. Similar application of isospin
symmetry and charge conjugation to the $\rho$ and $\sigma$
fragmentation functions that appear in the denominator of
Eq.~(\ref{asymmetry}) leads to a reduction in the number of
independent functions: $\hat{u}_i =\hat{d}_i=\hat{\bar u}_i=\hat{\bar
d}_i$ and $\hat{s}_i=\hat{\bar s}_i$ for $i=\{0,1\}$. 
For other systems the situation is more complicated due to the
relaxation of the Bose symmetry restriction. For example, for 
the $K\overline K$ system, $\delta \hat{q}^a_I=-\delta
\hat{\bar q}^a_{I}$ still holds, but $\delta\hat{u}_{I}$,
$\delta\hat{d}_{I}$, and $\delta\hat{s}_I$, are in general
independent. We also note
that application of the Schwartz inequality puts an upper bound on the
interference fragmentation function, $\delta\hat{q}_I^2\leq
4\hat{q}_0\hat{q}_1/3$ for each flavor.

\begin{figure}[h]
\begin{minipage}[h]{6.0in}
\epsfxsize=3.0in
\begin{center}
\mbox{\hspace*{-1cm}\epsfig{file=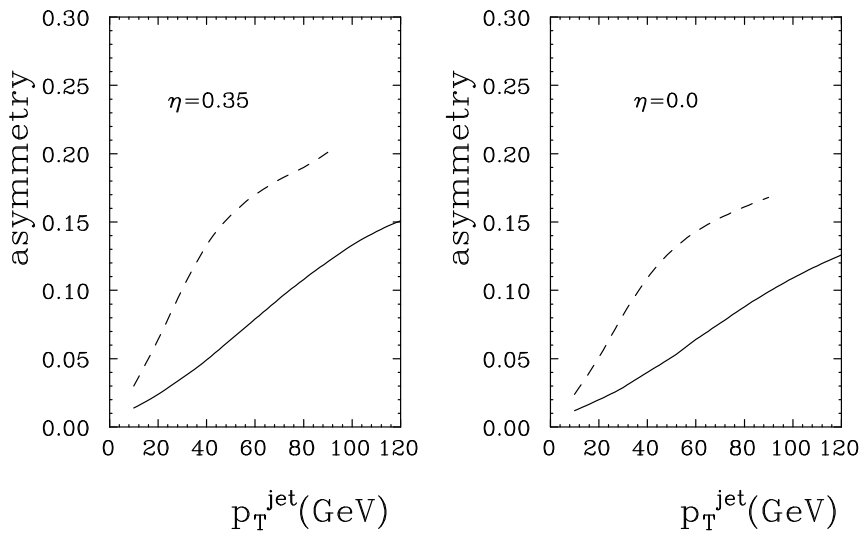,width=12.0truecm,angle=90}}
\end{center}
\end{minipage}
       \caption{
        The single spin symmetry as function of $p_T^{\rm jet}$ for two-pion
        production in $p p$ collision at $\protect\sqrt{s}=$500~GeV(solid)
        and $\protect\sqrt{s}=$200~GeV(dashes)
        (pseudo-rapidity $\eta=0.0$ and $\eta=0.35$ ).}
\label{fig3}
\end{figure}

The size of the asymmetry ${\cal A}_{\bot\top}$ critically 
depends upon the ratio of the $s-p$ interference fragmentation
function and $\rho$ and $\sigma$ fragmentation functions, which is 
unknown at present. In order to estimate the magnitude of
${\cal A}_{\bot\top}$, we saturate the  Schwartz inequality
and replace the interference fragmentation with its
upper bound, i.e. $\delta\hat{q}_{I}^2=
4\hat{q}_0\hat{q}_1/3$ for each flavor. Meanwhile, we assume the $\sigma$
and $\rho$ fragmentation functions are equal to each other.
Thus, the fragmentation function dependences
 cancel out in ${\cal A}_{\bot\top}$.
We also assume that the transversity $\delta q(x, Q^2)$ 
saturates the Soffer inequality\cite{soffer}: 
$2 |\delta q(x, Q^2)|= q(x, Q^2)+\Delta q(x, Q^2)$.
We use the polarized structure functions obtained by
Gehrmann and Stirling through next-to-leading order
analysis of experimental data\cite{gs}.
We also go to the region $m=0.83{\rm GeV}$, around which
the phase factor $|\sin\delta_0\sin\delta_1\sin(\delta_0-\delta_1)|$
is large (See Fig.~\ref{fig2}), and let $\cos\phi = 1$.
The asymmetry as function of $p_T^{\rm jet}$ at $\sqrt{s}=500 {\rm GeV}$ 
and $\sqrt{s}=200 {\rm GeV}$ for 
pseudo-rapidity $\eta=0.0$ and $\eta=0.35$ is shown in Fig.~\ref{fig3}, where
$p_T^{\rm jet}$ is the transverse momentum of the jet.
The size of asymmetry is about $12-15\%$ at $p_T^{\rm jet}=120 {\rm GeV}$ for
$\sqrt{s}=500 {\rm GeV}$ and about $17-20\%$ at $p_T^{\rm jet}=90 {\rm GeV}$ 
for $\sqrt{s}=200 {\rm GeV}$, which would be measurable at RHIC. 

A few comments can be made about our numerical results. 
Firstly, under the above approximations, the asymmetry is independent of
$z$. Of course, the experiment may not be able to determine $p_T^{\rm jet}$
--- the transverse momentum of the jet, so direct comparison
between our asymmetry and experimental data will require
event simulation. 
Secondly, because we don't know the sign of the unknown quantities yet,
we can not determine the sign of the asymmtery, the asymmetry shown in 
Fig.~\ref{fig3} should only be taken as its magnitude.
Finally, in order to estimate the asymmetry,
we have made very optimistic assumptions
about the novel interference fragmentation functions and
transversity distribution functions, so our estimates here
should be regarded as ``the high side''.

To summarize, we have studied the possibility of probing
the quark transversity distribution in the nucleon 
via two-meson semi-inclusive production of 
(only one beam) transversely polarized nucleon-nucleon collisions.
We obtained a single spin asymmetry that is sensitive
to the quark transversity and estimated its magnitude.

\vspace*{1cm}

I would like to thank Bob Jaffe for encouraging and enlightening
conversations relating to this subject.
I would also like to thank Xuemin Jin for helpful
discussions. I am also grateful to N. Saito and M. Baker 
for informations related to RHIC.



\begin{table}
\begin{center}
\begin{tabular}{|c|cc|}
\hline
Partonic process & Spin Average &
  Transversity Dependent\\
$ab\rightarrow cd$ &
  Cross Section$-\hat{\bar\sigma}_{ab}^{cd}$ & 
 Cross Section$-\delta\hat{\sigma}_{ab}^{cd}$\\
\hline
$qg\rightarrow qg$ &
 ${{\hat{s}^2+\hat{u}^2}\over {\hat{t}^2}}-{\frac 49}
{{\hat{s}^2+\hat{u}^2}\over {\hat{s}\hat{u}}}$
& ${{\hat{s}\hat{u}}\over {\hat{t}^2}}-{\frac 49}$\\[0.5cm]
$\bar{q}g\rightarrow \bar{q}g$ & 
 ${{\hat{s}^2+\hat{u}^2}\over {\hat{t}^2}}-{\frac 49}
{{\hat{s}^2+\hat{u}^2}\over {\hat{s}\hat{u}}}$
& ${{\hat{s}\hat{u}}\over {\hat{t}^2}}-{\frac 49}$\\[0.5cm]
$qq\rightarrow qq$ &
   ${\frac 49}\left({{\hat{s}^2+\hat{u}^2}
\over {\hat{t}^2}}+
{{\hat{s}^2+\hat{t}^2}\over {\hat{u}^2}}\right)-{\frac 8{27}}
{{\hat{s}^2}\over{\hat{u} \hat{t}} }$
&  ${\frac 4{27}}{{\hat{s}}\over{\hat{t}}}-{\frac 49}
{{\hat{s}\hat{u}}\over{\hat{t}^2}}$\\[0.5cm]
$qq'\rightarrow qq'$ &  ${\frac 49}{{\hat{s}^2+\hat{u}^2}\over{\hat{t}^2}}$
& $-{\frac 49}{{\hat{s}\hat{u}}\over{\hat{t}^2}}$\\[0.5cm]
$q\bar{q}\rightarrow q\bar{q}$
&  ${\frac 49}\left({{\hat{s}^2+\hat{u}^2}\over {\hat{t}^2}}+
{{\hat{u}^2+\hat{t}^2}\over {\hat{s}^2}}\right)-{\frac 8{27}}
{{\hat{u}^2}\over {\hat{s}\hat{t}}}$
&  ${\frac 8{27}}{{\hat{u}}\over{\hat{t}}}-{\frac 49}
{{\hat{s}\hat{u}}\over{\hat{t}^2}}$\\[0.5cm]
$q\bar{q}'\rightarrow q\bar{q}'$ &  
${\frac 49}{{\hat{s}^2+\hat{u}^2}\over{\hat{t}^2}}$
&  $-{\frac 49}{{\hat{s}\hat{u}}\over{\hat{t}^2}}$\\[0.5cm]
$gg\rightarrow q\bar{q}$ & ${\frac 16}{{\hat{t}^2+\hat{u}^2}
\over{\hat{t}\hat{u}}}-{\frac 38}{{\hat{t}^2+\hat{u}^2}
\over{\hat{s}^2}}$
& $---$\\[0.5cm]
\hline
\end{tabular}
\vspace*{1cm}
\hspace*{-2cm}\caption{ Partonic cross sections for 
$q_aq_b\rightarrow q_cq_d$ (only spin-average and transversity-dependent
parts are shown here).}
\end{center}
\label{table}
\end{table} 

\end{document}